\documentclass[aps,pre,reprint,groupedaddress]{revtex4-1}

\usepackage{epsfig}
\usepackage{graphicx}
\usepackage{epstopdf}
\usepackage{color}

\begin{document}

%\preprint{}

\title{Bond Percolation in Higher Dimensions}

\author{Eric I. Corwin}
\email[]{eric.corwin@gmail.com}
\affiliation{Department of Physics, University of Oregon, Eugene, Oregon 97403}

\author{Robin Stinchcombe}
\email[]{r.stinchcombe1@physics.ox.ac.uk}
\affiliation{Rudolf Peierls Centre for Theoretical Physics,
University of Oxford, 1 Keble Road, Oxford OX1 3NP, U.K}

\author{M.F. Thorpe}
\email[]{mft@asu.edu}
\affiliation{Department of Physics,
Arizona State University, Tempe, AZ 85287-1604, U.S.A}
\affiliation{Rudolf Peierls Centre for Theoretical Physics,
University of Oxford, 1 Keble Road, Oxford OX1 3NP, U.K}

\begin{abstract}

We collect together results for bond percolation on various lattices from two to fourteen dimensions which, in the limit of large dimension $d$ or number of neighbors $z$, smoothly approach a randomly diluted Erd\H{o}s-R\'{e}nyi graph.  We include new results on bond diluted hyper-sphere packs in up to nine dimensions, which show the mean coordination, excess kurtosis and skewness evolving smoothly with dimension towards the Erd\H{o}s-R\'{e}nyi limit.

%Using results for bond percolation on various lattices from 2 to 14 dimensions, we show for the first time that in the limit of large dimension $d$ and/or large number of neighbors $z$, a randomly diluted Erd\H{o}s-R\'{e}nyi graph is approached smoothly. This conclusion is reinforced with new results on bond diluted hyper-sphere packs which show how the mean coordination, excess kurtosis and skewness evolve with dimension to the Erd\H{o}s-R\'{e}nyi limit. It is demonstrated that this result occurs because loops become irrelevant in percolation in high dimensions.
 
\end{abstract}

\pacs{xxx,xxx,xxx}

\maketitle

\section{Introduction}

\begin{figure}[h]
\centering
\includegraphics[scale=0.9]{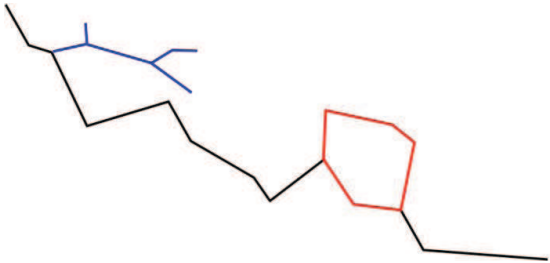}
\caption{Showing a connected path in black across part of a sample with \textit{redundancy} via the red loop which is over-constrained with one redundant bond and \textit{irrelevancy} via the blue region that contains dangling ends that are not involved in percolation.}
\label{fig:fig1}
\end{figure}

Percolation theory~\cite{stauffer_introduction_1994, essam_percolation_1980} asks if there is a connected path across a system. Examples are water percolating through ground coffee beans and forest fires spreading from tree to tree. In order to have a control parameter, percolation is often studied on lattices of dimension $d$ greater than 1, where percolation disappears when the random removal of bonds has decreased the bond concentration $p$ to a critical value $p_c$. The disappearance of percolation is well studied and is a second order phase transition with a set of critical exponents that obey scaling laws~\cite{stauffer_introduction_1994,essam_percolation_1980}. Our interest here is in the values of $p_c$ studied over diverse geometries.  Although this is an old subject, interest continues, including in higher dimensions, where rigorous bounds on $p_c$ have recently been established~\cite{torquato_effect_2013}.  We denote by $z$ the number of initial bonds at any site of a particular regular lattice, (e.g. triangular net, simple cubic, etc.) before bond dilution occurs ($p=1$). It is convenient to define the mean coordination $\langle r \rangle$ at the percolation point as
%ERIC NOTE: I don't understand what Mike is getting at here
%
\begin{equation}
\langle r \rangle = z p_c,
\label{equation1}
\end{equation}
which facilitates the comparison between various lattices in various dimensions as the mean coordination $\langle r \rangle$ at percolation varies much less than $p_c$ itself. A very simple argument suggests that $\langle r \rangle = 2$ at the transition as each site must have one bond entering and one bond leaving to form a connected pathway. While this is the most efficient scenario, it does not happen quite this way in a random system for two reasons. First, there is \textit{redundancy} where there is more than one connection between two points, leading to a loop. Loops push the mean coordination $\langle r \rangle$ above 2, because at least two sites with coordination 3 must be involved in forming a loop. Second, there is \textit{irrelevancy} where dead-ends and isolated regions are formed that would not carry a current if the bonds were wires in a conducting network.  Irrelevancy pushes the mean coordination $\langle r \rangle$ below 2, as some sites are singly coordinated. Both these situations are illustrated in Figure~\ref{fig:fig1}. We will see that there is a tendency for the effects of redundancy and irrelevancy to cancel making $\langle r \rangle=2$ a not unreasonable starting approximation for low dimensions $d$ and/or low initial coordination $z$.  However, for very high dimension $d$ or initial coordination $z$, the mean coordination number $\langle r \rangle$ approaches unity because of the preponderance of dangling bonds.

The result $\langle r \rangle = 2$ at the transition can also be derived by Maxwell type constraint counting~\cite{maxwell_calculation_1864} of the number of floppy modes~\cite{thorpe_continuous_1983,thorpe_flexibility_2009} or residual degrees of freedom $f$ in the system. Connectivity percolation, which is the subject of this note, can be regarded as a special case of a larger class of problems where instead of having a single degree of freedom per site there are $g$ degrees of freedom.  An example would be vector displacements in two dimensions, where $g = 2$ .  For $g \geq 2$ such problems are usually referred to as \textit{rigidity percolation}~\cite{thorpe_continuous_1983,thorpe_flexibility_2009}.

Maxwell constraint counting~\cite{maxwell_calculation_1864} is more usually employed in problems involving rigidity, but can also be applied to connectivity percolation problems as a special case, with $g = 1$.  More generally, there are $g$ degrees of freedom associated with each site and $z$ constraints are present (the number of bonds at each site is assumed to be exactly $z$ everywhere initially) with probability $p$, so that
\begin{equation}
f=g-zp/2,
\label{equation2}
\end{equation}
which goes to zero at $p_c = 2g/z$, and hence gives the result $\langle r \rangle = 2g$ at percolation. Note that the number of floppy modes is not exactly zero at the transition as fluctuations in local coordination number allow for local redundancy and irrelevancy, but nevertheless it has been shown that the number of floppy modes at the transition is extremely low~\cite{jacobs_generic_1996}, making  $\langle r \rangle = 2g$ an unusually accurate approximation for $g \geq 2$ (typically within one percent).  For example, in the case of rigidity percolation of a triangular net under bond dilution Maxwell counting gives a result of $\langle r \rangle = 4$ while numerical simulations \cite{jacobs_generic_1996} find $\langle r \rangle = 3.961 \pm 0.002$, which is very close to, but clearly less than, 4.  However, the constraint counting result $\langle r \rangle = 2$ for connectivity percolation gets worse in higher dimensions in which $\langle r \rangle = 1$ is reached.  Nevertheless, $\langle r \rangle$ is a more useful variable than $p_c$ as it changes much less rapidly with dimension, and we will focus on it here.

\begin{figure}[b]
\centering
\includegraphics[scale=0.25]{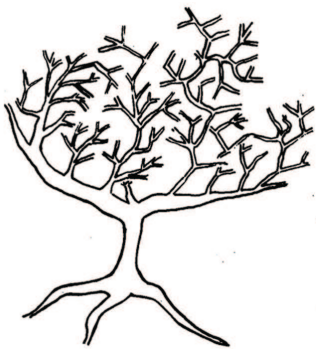}
\caption{Showing a tree or Bethe lattice, reproduced from reference~\cite{thorpe_bethe_1982}}
\label{fig:fig2}
\end{figure}

\section{Bethe Lattice}
\begin{figure}[h]
\centering
\includegraphics[scale=0.3]{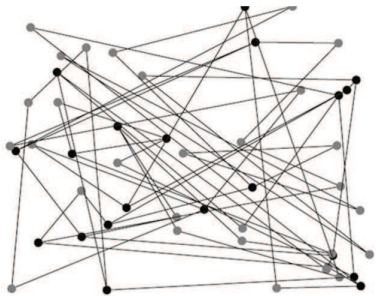}
\caption{Showing a bond diluted finite Erd\H{o}s-R\'{e}nyi graph, where before dilution every node was connected to every other node.}
\label{fig:fig3}
\end{figure}

A useful universal guideline is provided by the Bethe lattice which is a tree-like network that contains no loops as illustrated in Figure~\ref{fig:fig2}~\cite{ fisher_cluster_1961, thorpe_bethe_1982}. If each node of the tree is $z$ coordinated before dilution, then for a connected path there must be one way in from a previous layer, and one of the remaining $z - 1$ ways out must be occupied, so that $p_c = 1/(z-1)$; a result which can be rigorously found~\cite{fisher_cluster_1961}. Hence the mean coordination $\langle r \rangle$ at percolation is given by
\begin{equation}
\langle r \rangle = zp_c =  z/(z-1).
\label{equation3}
\end{equation}
Of course not all sites have exactly this coordination as there is a binomial distribution of local coordination numbers due to the random dilution; so the probability of a site having $r$ bonds present out of a total of $z$ possible is given by
\begin{equation}
%P(r)  =  \sum\limits_{r=0}^z {^z}C_r   p^r (1-p)^{z-r}
P(r)  =  \sum\limits_{r=0}^z {z \choose r}   p^r (1-p)^{z-r}
\label{equation4}
\end{equation}
and hence the $n^{th}$ moment $\langle r^n \rangle$ is given by
\begin{equation}
\langle r^n \rangle = \sum\limits_{r=0}^z r^n P(r)
\label{equation5}
\end{equation}
leading to the mean coordination
\begin{equation}
\langle r \rangle =zp
\label{equation6}
\end{equation}
and the square of the width $\Delta r$ given by
\begin{equation}
(\Delta r)^2 = \langle r^2 \rangle  - {\langle r \rangle}^2 =zp(1-p)
\label{equation7}
\end{equation}
This expression for the width is quite general for any network with fixed initial coordination $z$ at every site, upon random bond dilution. For the Bethe lattices at the percolation threshold, this width becomes
\begin{equation}
 \Delta r =  \frac{\sqrt{z(z-2)}}{(z-1)}
\label{equation8}
\end{equation}
A particularly interesting limit is large $z \rightarrow\infty$ where we obtain what we will refer to as the Erd\H{o}s-R\'{e}nyi limit; reached when percolation occurs upon bond dilution in a graph that initially has $N$ nodes, each one connected to every other node~\cite{erdos_evolution_1960} as $N \rightarrow \infty$.  In this limit
\begin{equation}
\langle r \rangle = \Delta r = 1.
\label{equation9}
\end{equation}
This is the limit of a large graph of nodes, where every node is connected to every other node with probability $p$. In the limit that the number of nodes goes to infinity, the chance of finding a loop becomes infinitesimally small and hence the large $z$ Bethe lattice result is obtained. An example of a finite Erd\H{o}s-R\'{e}nyi graph~\cite{erdos_evolution_1960} is shown in Figure~(\ref{fig:fig3}).

\section{A Universal plot}
\begin{figure}[h]
\centering
\includegraphics [scale=0.9]{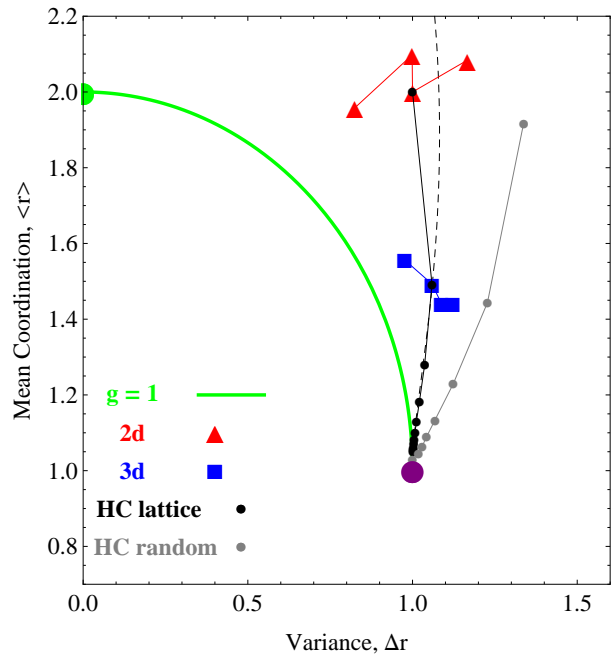}
\caption{Showing results for the mean coordination against the variance for 2d lattices (red), 3d lattices (blue), hypercubic lattices (black) and random hypersphere packings (gray), at the percolation threshold. The straight lines joining adjacent points are only for guidance of the eye.  The green line is the Bethe lattice result with the isostatic point at (0,2) and the Erd\H{o}s-R\'{e}nyi result (1,1) shown as the large purple dot.  The dashed line shows the result of a $1/(z-1)$ expansion \cite{gaunt_bond_1978} given in equation (\ref{equationGauntRuskin}) for hypercubic lattices}
\label{fig:fig4}
\end{figure}

Using what we have jotted down in the previous paragraph, it is convenient to combine all results for bond percolation on various lattices as a plot of the mean coordination $\langle r \rangle$ against the width of the distribution, or variance, $\Delta r$ which is shown in Figure~(\ref{fig:fig4}).

The results for the $2d$, $3d$ and hypercubic lattices are conveniently summarized with original references in Wikipedia~\cite{_percolation_2013}. The two dimensional results, shown in red in Figure~(\ref{fig:fig4}), are from left to right, following the thin red line, honeycomb, kagome, square net and triangular net. The general trend is higher initial coordination $z$ to the right going to lower initial coordination $z$ to the left, which tends to the isostatic point shown at (0,2). The point for the kagome lies above the point for the square net in the center and gives an idea of the (modest) effect of the detailed lattice structure as both have sites with four neighbors initially. Nevertheless the overall trend that the red points get closer to the isostatic point as the initial coordination $z$ is decreased is clear.

The three dimensional results, shown in blue, are from left to right, following the thin blue line, diamond, simple cubic, body centered cubic and face centered cubic; with the latter two close together but following the general trend with higher initial coordination $z$ to the right going to lower initial coordination $z$ to the left, which again tends to the isostatic point at (0,2). Also included in Figure~(\ref{fig:fig4}) are the results for bond diluted hypercubic lattices from $d=2$ up to $d=13$ where the mean coordination $\langle r \rangle$ is obtained from (\ref{equation1}) and the variance from (\ref{equation7}).

The results for diluted non-crystalline hypersphere packings are new and were obtained from computer simulations of jammed configurations of $N = 262144$ monodisperse particles (in $2d$ a 50-50 mixture of bidisperse particles with size ratio 1.4:1 was used to avoid crystallization) as described in reference~\cite{charbonneau_universal_2012}. The particles interact with a harmonic contact potential defined as

\begin{equation}
V(r) =  \epsilon \left(\sigma - r\right)^2 \Theta \left(\sigma - r\right)
\label{equation9a}
\end{equation}
where $\sigma$ is the particle radius, ${\epsilon}$ the energy scale of the potential, and $r$ the distance between particles. Energy is minimized at a given packing fraction via either a conjugate gradient~\cite{hestenes_methods_1952} or fast inertial relaxation engine (FIRE)~\cite{bitzek_structural_2006}  minimization technique. Starting from a random configuration at a density well above jamming and given two values of packing fraction that bracket the jamming transition density the jamming point is found via a golden mean bisection search. Jamming is identified as the packing fraction corresponding to the onset of non-zero energy as derived from the potential (\ref{equation9a}). For the purposes of percolation studies, two hyperspheres are said to be connected neighbors if there is a non-zero overlap between them.  We find that the values of $\langle r \rangle$ and $\Delta r$ are rather insensitive to a (small) distance from the jamming transition.  

Note that both sets of high-dimensional results, for bond percolation in hypercubic lattices and random hypersphere packings approach the Erd\H{o}s-R\'{e}nyi limit, as can be seen from Figure (\ref{fig:fig4}).  The highest dimension explored of $d=13$ for hypercubic lattices and $d=9$ for random hypersphere packings are already very close to the point (1,1). This is because loops become less important in higher dimensions, discussed next.

%UNCLEAR TO ME - ERIC CORWIN%%%%%%%%%%%%

If we consider the mean and variance of the percolation variable $r$ then the Erd\H{o}s-R\'{e}nyi limit is (from Eqns. \ref{equation3} - \ref{equation9}) the same as the $z$-going-to-infinity limit of the Bethe lattice (tree) result. The tree is, in turn, the loopless limit of a general lattice; and, from simple geometric path counting considerations, the loopless limit is the large $z$, and equivalently the large $d$, limit of a general lattice.  The probability of two sites being joined by a graph with $n$ links will be proportional to $p_c^n$.  Now consider all graphs with $n$ steps. For the ``trees'' we have $r=n/2$, and for all other graphs with a partial or full loop $r>n/2$. The key observation is that in high dimensions, $p$ goes like $1/d$, as can be seen for the Bethe lattice in Equation (8).  For example, hypercubic lattices have $z=2d$  and for random packings $z$ is even larger. Therefore  as $d$ goes to infinity those diagrams with $r=n/2$ overwhelmingly dominate and hence only the trees contribute, and the Erd\H{o}s-R\'{e}nyi limit is reached. While this is not a formal proof, it demonstrates the plausibility of the result, and should form the basis for a formal mathematical proof.

%%%%%%%%%%%%%%%%%%%%%%%%%%%%%%%%%%%%%

For completeness, we include the results of Gaunt and Ruskin~\cite{gaunt_bond_1978} who performed a $1/(z-1)$ expansion for bond percolation on bond diluted hypercubic lattices where $z=2d$ and found that percolation occurs at
\begin{equation}
p_c = \sigma[1 + \frac{5}{2}\sigma^2 + \frac{15}{2}\sigma^3 + 57\sigma^4 + ... ]
\label{equationGauntRuskin}
\end{equation}
where $\sigma = (z-1)^{-1}$. Note that the leading term is the Bethe lattice result. 
%From equation~(\ref{equationGauntRuskin}) as well as equations~(\ref{equation1}) and~(\ref{equation7}), 
We include this result in Figure~(\ref{fig:fig4}) as a dashed line, which is seen to be very close indeed to the results of numerical simulations (black dots) for hypercubic lattices with $d \ge 3$, then deviating at $d=2$ for the square lattice.

\begin{figure}[h]
\centering
\includegraphics [scale=0.95]{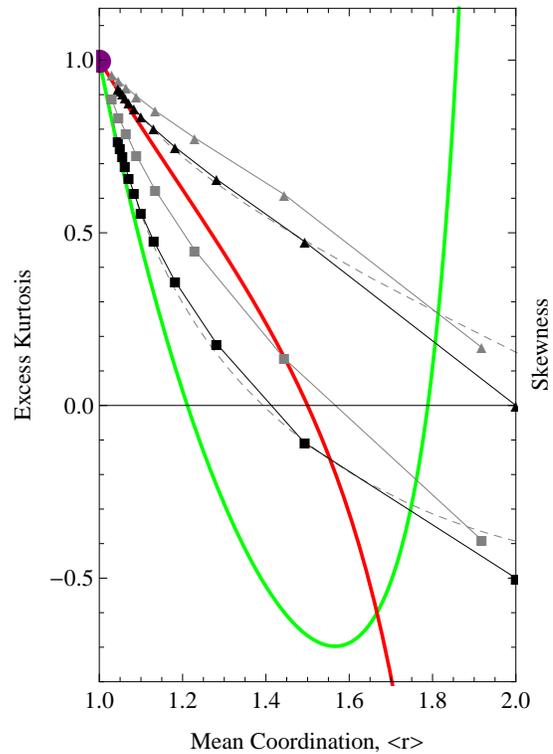}
\caption{Showing the skewness (red line) and the excess kurtosis (green line) as a function of the mean coordination $\langle r \rangle$ for Bethe lattices at the percolation threshold. Also shown are the skewness (triangles) and excess kurtosis (squares) for hypercubic lattices as gray symbols and random hypersphere packings as black symbols. The straight lines joining adjacent symbols are guides to the eye.  The Erd\H{o}s-R\'{e}nyi result (1,1) is shown as the large purple dot.  The dashed line shows the result of a $1/(z-1)$ expansion \cite{gaunt_bond_1978} given in equation (\ref{equationGauntRuskin})}
\label{fig:fig5}
\end{figure}

Another convenient way to monitor the approach of dilute hypercubic lattices to the Erd\H{o}s-R\'{e}nyi limit, is to track the {\it skewness} $\gamma_1$ and {\it excess kurtosis} $\gamma_2$  which respectively monitor the evolution of the asymmetry and the deviation from Gaussian behavior of the distribution of contacts (for a Gaussian distribution $\gamma_1 = \gamma_2 = 0$). These are defined in terms of the moments of the distribution as
\begin{equation}
\gamma_1 = \frac{\langle \left( r  - \langle r \rangle \right)^3 \rangle}{{\langle \left( r - \langle r \rangle \right)^2 \rangle}^{3/2}},
\label{equation10a}
\end{equation}
\begin{equation}
\gamma_2 = \frac{\langle \left( r  - \langle r \rangle \right)^4 \rangle}{{\langle \left( r  - \langle r \rangle \right)^2 \rangle}^{2}} - 3
\label{equation11a}
\end{equation}
For Bethe lattices they take the values
\begin{equation}
\gamma_1 = \frac{(3-2\langle r \rangle)}{\sqrt{\langle r \rangle\left(2-\langle r \rangle\right)}}
\label{equation10}
\end{equation}
\begin{equation}
\gamma_2 =  \frac{1}{\langle r \rangle(2-\langle r \rangle)} - \frac{6\left(\langle r \rangle-1\right)}{\langle r \rangle}
\label{equation11}
\end{equation}
These are plotted as the solid lines in Figure~(\ref{fig:fig5}). In the limit of a Bethe lattice with large $z$, the distribution of coordination number becomes a Poisson distribution with $p(r) = e^{-1}/r!$ and thus $\langle r \rangle  = \Delta r = \gamma_1 = \gamma_2 = 1$. Note that for the Bethe lattice, the skewness  goes through zero at $\langle r \rangle$ = 3/2 which corresponds to $z$ =3, and the excess kurtosis goes through zeros at $\langle r \rangle = \left(9 \pm \sqrt{3}\right)\/6$ = 1.211 and 1.789 which corresponds to $z = 4 \pm \sqrt{3} $ = 2.227 and 6.928 respectively.

The skewness and the excess kurtosis for the hypercubic lattices can be obtained for the known values of $p_c$ from reference~\cite{_percolation_2013} and using equations~(\ref{equation4}) and~(\ref{equation5}) respectively. For the binomial distribution, the skewness is
\begin{equation}
\gamma_1 = \frac{1-2p}{\sqrt{zp\left(1-p\right)}}
\label{equation10b}
\end{equation}
and the excess kurtosis is
\begin{equation}
\gamma_2 = \frac{1}{zp\left(1-p\right)} - \frac{6}{z}
\label{equation10c}
\end{equation}
and these are also plotted at the percolation threshold in Figure (\ref{fig:fig5}), which shows how they approach the Erd\H{o}s-R\'{e}nyi limit in high dimensions, providing further evidence of the relative unimportance of loops in connectivity percolation in higher dimensions.  Results for the skewness and excess kurtosis can also be obtained from the expansion~\cite{gaunt_bond_1978} given in equation (\ref{equationGauntRuskin}), coupled with equations~(\ref{equation10}) and (\ref{equation11}), and are shown as the dashed lines in Figure (\ref{fig:fig5}).  Also shown in Figure~\ref{fig:fig5} are directly computed results for the skewness and excess kurtosis for bond-diluted random hypersphere packs at the percolation threshold in higher dimensions.  Again a similar trend towards the Erd\H{o}s-R\'{e}nyi limit in high dimensions is very apparent.  All results for bond-diluted hypersphere packings at the percolation threshold are tabulated in Table (\ref{table:hypersphere}).

\begin{table}
 \begin{center}
 \setlength{\tabcolsep}{5pt}

\begin{tabular}{l||l|l|l|l}\hline
    $d$ & $\langle r \rangle$ & $\Delta r$ & Skewness & Excess Kurtosis\\\hline \hline
    2 &   1.9174 &   1.3373 &   0.1687 &  -0.3890 \\\hline
    3 &   1.4435 &   1.2274 &   0.6113 &   0.1368 \\\hline
    4 &   1.2289 &   1.1234 &   0.7749 &   0.4499 \\\hline
    5 &   1.1338 &   1.0682 &   0.8535 &   0.6242 \\\hline
    6 &   1.0890 &   1.0423 &   0.8954 &   0.7244 \\\hline
    7 &   1.0642 &   1.0292 &   0.9214 &   0.7891 \\\hline
    8 &   1.0459 &   1.0181 &   0.9397 &   0.8355 \\\hline
    9 &   1.0294 &   1.0104 &   0.9596 &   0.8877 \\\hline
 \end{tabular}
 \end{center}

\caption{Tabulated values for $\langle r \rangle$, $\Delta r$, Skewness, and Excess Kurtosis for random hypersphere packings of $N = 262144$ particles in dimensions $d=2 - 9$.  Note that all packings are constructed with monodisperse spheres except for $d=2$ for which a 50-50 mixture of bidisperse particles with size ratio 1.4:1 is used.}
\label{table:hypersphere}
\end{table} 

These kinds of  argument extend from percolation to a range of other processes. Among them are other $q$ state Potts models
(the $q \rightarrow 1$ limit is  bond percolation~\cite{fortuin_random-cluster_1972}), which includes the Ising model ($q = 2$). This was perhaps the first system for which small $1/z$ was systematically exploited by Brout and Englert~\cite{brout_phase_1966,englert_linked_1963}. The limit $1/z \rightarrow 0$ gives mean field theory, associated with the tree graphs of the linked cluster many-body theory. This is the starting point for a $1/z$ expansion involving graphs with increasing numbers of loops, which account for the fluctuation effects absent from mean field theory. It is interesting to note that similar arguments to those given here were previously given by Brout~\cite{brout_phase_1966} who exploited the link between trees and mean field theory for the Ising model, using large $z$, where the factor ${\left(J\/k_B T\right)}^n$ in an $n^{th}$ order graph being analogous to the ${p_c}^n$ here. In the Ising model $J$ is the exchange interaction between spins and $T$ is the temperature.

The related role of higher dimensions reducing fluctuations is of course well known in such contexts~\cite{gaunt_bond_1978, torquato_effect_2013}, as is its role in reducing the probability of returns to the origin (loops) in random walks and related {\it dynamic} processes. These aspects suggest future work exploiting the approach used here for other processes.

\section{Conclusions}
We have shown that all bond dilution results have universal features so that results for various lattices in various dimensions can be displayed on a single plot and these results approach the Erd\H{o}s-R\'{e}nyi limit in high dimensions. The Erd\H{o}s-R\'{e}nyi limit  is when percolation occurs upon bond dilution in a graph that initially has $N$ nodes each one connected to every other one~\cite{erdos_evolution_1960} as $N \rightarrow \infty$. It is also shown here that the mean coordination at percolation $\langle r \rangle$ is often a more useful universal parameter than the percolation concentration itself $p_c$.

\section{Acknowledgments} We should like to thank the US National Science Foundation for support under Career Award DMR-1255370 (EIC) and DMR-0703973 (MFT) and by a Major Research Instrumentation grant, Office of Cyber Infrastructure, ``MRI-R2: Acquisition of an Applied Computational Instrument for Scientific Synthesis (ACISS),'' Grant No. OCI-0960354.

\bibliographystyle{unsrt}
\bibliography{UniversalBondPercolation.bib}

\begin{thebibliography}{10}

\bibitem{stauffer_introduction_1994}
Dietrich Stauffer and Ammon Aharony.
\newblock {\em Introduction To Percolation Theory}.
\newblock {CRC} Press, July 1994.

\bibitem{essam_percolation_1980}
J.~W. Essam.
\newblock Percolation theory.
\newblock {\em Reports on Progress in Physics}, 43(7):833, July 1980.

\bibitem{torquato_effect_2013}
S.~Torquato and Y.~Jiao.
\newblock Effect of dimensionality on the percolation thresholds of various
  d-dimensional lattices.
\newblock {\em Physical Review E}, 87(3):032149, March 2013.

\bibitem{maxwell_calculation_1864}
{JC} Maxwell.
\newblock On the calculation of the quilibrium and stiffness of frames.
\newblock {\em Philosophical Magazine}, 27:294--299, 1864.

\bibitem{thorpe_continuous_1983}
{M.F.} Thorpe.
\newblock Continuous deformations in random networks.
\newblock {\em Journal of Non-Crystalline Solids}, 57(3):355--370, September
  1983.

\bibitem{thorpe_flexibility_2009}
{MF} Thorpe.
\newblock Flexibility and mobility in networks encyclopedia of complexity and
  systems science.
\newblock In {RA} Meyers, editor, {\em Encyclopedia of Complexity and Systems
  Science}, volume~5, pages 6013--6024. Springer, New York, 2009.

\bibitem{jacobs_generic_1996}
D.~J. Jacobs and M.~F. Thorpe.
\newblock Generic rigidity percolation in two dimensions.
\newblock {\em Physical Review E}, 53(4):3682--3693, April 1996.

\bibitem{thorpe_bethe_1982}
{MF} Thorpe and {MF} Thorpe.
\newblock Bethe lattices.
\newblock In {\em Excitation in Disordered Systems}, {NATO} Advanced Study
  Institute Series B78, pages 85--107. Plenum Press, New York, 1982.

\bibitem{fisher_cluster_1961}
Michael~E. Fisher and John~W. Essam.
\newblock Some cluster size and percolation problems.
\newblock {\em Journal of Mathematical Physics}, 2(4):609--619, July 1961.

\bibitem{_percolation_2013}
Percolation threshold, April 2013.
\newblock Page Version {ID:} 548389631.

\bibitem{charbonneau_universal_2012}
Patrick Charbonneau, Eric~I. Corwin, Giorgio Parisi, and Francesco Zamponi.
\newblock Universal microstructure and mechanical stability of jammed packings.
\newblock {\em Physical Review Letters}, 109(20):205501, November 2012.

\bibitem{hestenes_methods_1952}
Magnus~R Hestenes and Eduard Stiefel.
\newblock Methods of conjugate gradients for solving linear systems1.
\newblock {\em Journal of Research of the National Bureau of Standards}, 49(6),
  1952.

\bibitem{bitzek_structural_2006}
Erik Bitzek, Pekka Koskinen, Franz Gähler, Michael Moseler, and Peter Gumbsch.
\newblock Structural relaxation made simple.
\newblock {\em Physical Review Letters}, 97(17):170201, October 2006.

\bibitem{gaunt_bond_1978}
D.~S. Gaunt and H.~Ruskin.
\newblock Bond percolation processes in d dimensions.
\newblock {\em Journal of Physics A: Mathematical and General}, 11(7):1369,
  July 1978.

\bibitem{fortuin_random-cluster_1972}
{C.M.} Fortuin and {P.W.} Kasteleyn.
\newblock On the random-cluster model: I. introduction and relation to other
  models.
\newblock {\em Physica}, 57(4):536--564, February 1972.

\bibitem{brout_phase_1966}
Robert~H. Brout.
\newblock Phase transitions.
\newblock {\em American Journal of Physics}, 34(9):830, 1966.
\newblock see especially Chapter 2, section 5.

\bibitem{englert_linked_1963}
F.~Englert.
\newblock Linked cluster expansions in the statistical theory of
  ferromagnetism.
\newblock {\em Physical Review}, 129(2):567--577, January 1963.

\end{thebibliography}

\end{document}